# Self-Organizing Mobility Robustness Optimization in LTE Networks with eICIC

DRAFT V5.0

Carl Weaver

Pantelis Monogioudis

Incubation & Innovation Team, Wireless CTO

Alcatel-Lucent

*Abstract*

We address the problem of Mobility Robustness Optimization (MRO) and describe centralized Self Organizing Network (SON) solutions that can optimize connected-mode mobility Key Performance Indicators (KPIs). Our solution extends the earlier work of eICIC parameter optimization [7], to heterogeneous networks with mobility, and outline methods of progressive complexity that optimize the Retaining/Offloading Bias which are macro/pico views of Cell Individual Offset parameters. Simulation results under real LTE network deployment assumptions of a US metropolitan area demonstrate the effects of such solutions on the mobility KPIs. To our knowledge, this solution is the first that demonstrates the joint optimization of eICIC and MRO.

# 1 Introduction

Mobility management in LTE includes two types of procedures: idle mode [1] and connected mode [2]. Idle mode[1] procedures include selection and re-selection of the best serving cell, maintenance of tracking area registration and transitioning to connected mode. Connected mode (handover) procedures in LTE comply with [2] and have been covered in [6]. The handover procedure, as shown in Figure 1, transitions the connection of the UE between a source cell, and a target cell. The UE when it detects that radio quality conditions satisfy the so called “A3 event” entering condition, further outlined below, initiates the handover procedure by sending a Measurement Report (MR) in the Physical Uplink Control/Shared Channel (PUCCH/PUSCH) channel to the source cell. The source cell evaluates the report and makes a vendor-proprietary decision to request a handover from the target cell. It ultimately responds with a Handover Command carried in the PDSCH channel. The typical time between the reception of the measurement report and the transmission of the Handover Command is short enough that it only makes a small difference in the overall failure rate of handovers. A Handover Complete command is sent by the UE to signal the successful cell change.

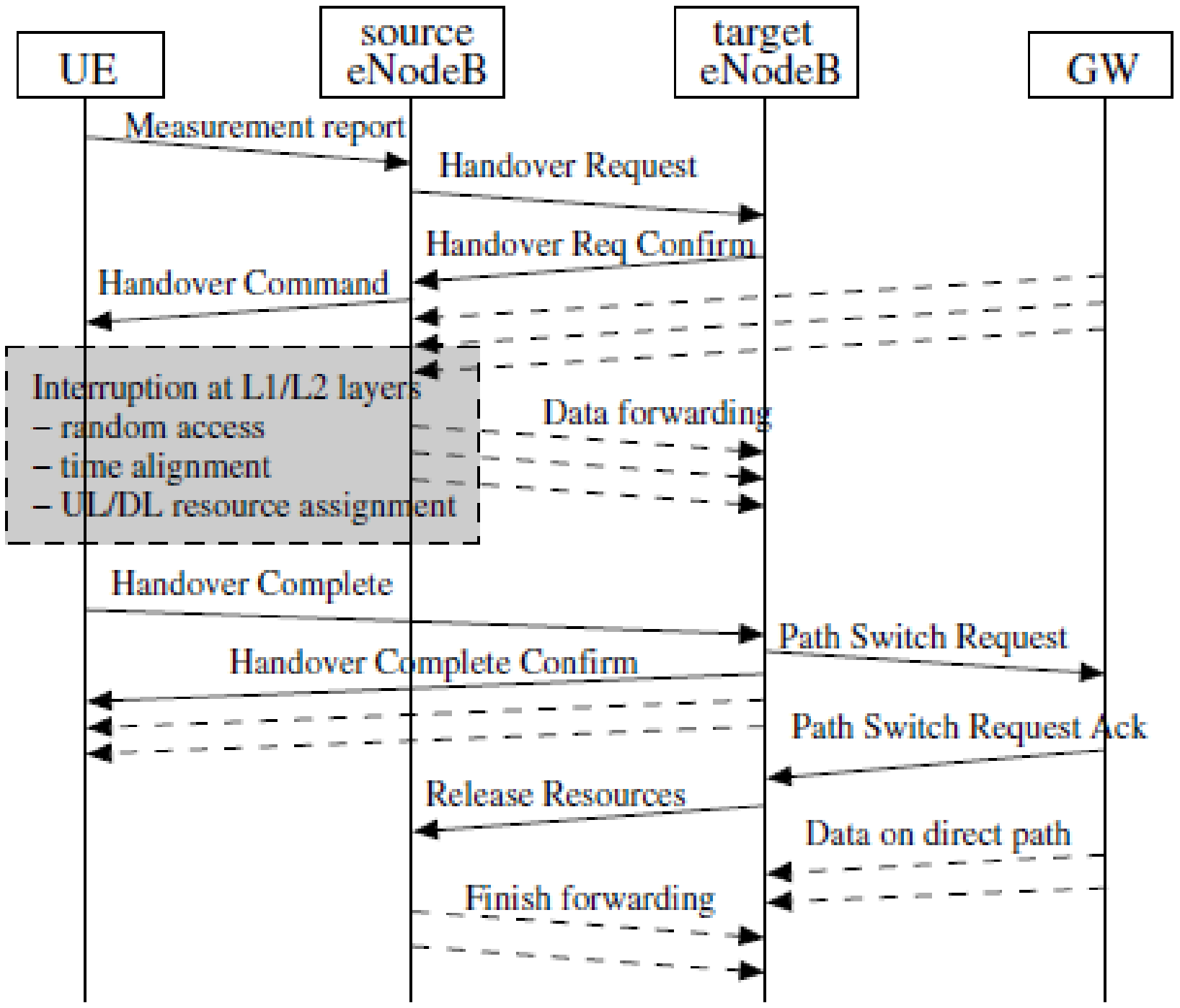

Figure 1: LTE HO Message Sequence Chart

[1] Idle mode parameters optimization are important, as they affect the idle-mode power consumption of the UE and the success rate of radio resource control (RRC) connection attempts. They deserve a dedicated treatment and are outside the scope of this paper.

The KPIs relevant to handover are shown in Table 1. The metric most affecting the Quality of Experience (QoE) for end users is the Drop Call Rate which depends on both handover failure probability (HOPF) and handover rate (HOR). Other metrics cover ping-pong conditions, short time-of-stay on a cell, and the race condition that is explained shortly.

The Mobility Robustness Optimization (MRO) work in 3GPP aims to optimize the parameters of Table 2. 3GPP MRO [4] classifies handover failures as Late, Early, or Early/Wrong Cell and advocates minimizing those failures as well as ping-pong and unwanted handovers by controlling these parameters. Experience in intra-carrier handovers in the field, has shown that for vehicular mobility cases, the majority of handover failures are caused by decoding failure of the PDCCH UL Grant message needed by the UE to send the MR in the uplink. This is despite that the UL Grant is typically repeated many times until the declaration of a Radio Link Failure (RLF) event [3]. Improvements on the PDCCH channel design are currently ongoing in 3GPP, but for the short term the only possible solution to avoid RLF events is the optimization of the MR triggering event that initiates the handover procedure. According to [4], the mobility parameters that can be optimized in connected mode are: Hysteresis, Time to Trigger (denoted as TTT or $t_T$ in this paper) and Cell Individual Offsets (CIOs)). Additional parameters may also be included in vendor-specific enhancements.

The condition that triggers a handover is the A3 event entering condition [2] :

$$M_n + O_{cn} - H_{ys} > M_p + O_{cp} + O_{ff} \qquad (1)$$

where $M_p$ is primary serving cell Reference Signal Received Power (RSRP) measurement and $M_n$ is a neighbor cell RSRP measurement. RSRP is measured on a broadcast channel, having pilot channel semantics, called the Cell Reference Signal (CRS) that is broadcast from each cell on specific resource elements. Other parameters in (1) are defined in Table 2, but briefly, $O_{cn}$ (a vector over neighbors) and $O_{cp}$ (a scalar) are CIO values and $O_{ff}$ and $H_{ys}$ are effective hysteresis for the entering condition.

In heterogeneous networks (HTN), further complications arise due to interference. The Enhanced Inter-Cell Interference Coordination feature (eICIC) was introduced to reduce interference caused to pico UEs, from neighboring overlaid macro cells, by nulling the power transmitted over a fraction of macro cell's time and frequency resources (sub-frames). Such sub-frames are known as Almost Blank Sub-frames (ABSs). During ABS, the Physical Dedicated Control/Data Channels (PDCCH/PDSCH) are not transmitted and the pico can then increase its range attempting to take as many users as optimally determined from the nearby macros. The increased spatial area is called Cell Range Extension (CRE) region. The primary means of range expansion is biasing the UEs RSRP measurements, using the CIO parameters, such that the pico appears stronger than it actually is. This causes increased A3 events from the macros towards the pico and therefore increased traffic offloading. Obviously, the artificial range extension has its limits - in many instances limits are imposed by CRS that are not blanked during the ABS sub-frames and interference between PCH/SCH not in ABS sub-frames as well as the actual achievable rate of PDSCH channel which decreases as the CRE region is increased.

It is important to highlight the coupling of ABS and bias [7] when the network-wide utility is maximized. It is also important to realize that in actual networks we can have many-to-many relationships: many picos interacting with many macros. Designs that ignore such parameter couplings and handover performance impacts of those couplings will be suboptimal. As we shall see later, although the use of CIO may seem intuitively simple, independent per-cell setting of CIO parameters can cause handover race conditions that are particularly problematic for services such as VoIP.

**Table 1: Key Performance Indicators in LTE Handover Procedures**

| KPI | Target KPI | Description |
|---|---|---|
| **Drop Call Rate** | $10^{-4}/\text{s}$ | Radio Link Failures per second in connected mode, assuming idle mode performance is approximately synced; The target in VoLTE is equivalent to UMTS or IS2000 3G1x CS Voice 1%/100s without RLF re-establishment recovery; the drop rate depends on both the HO Rate (HOR) and Handover Failure Probability (HOFP) |
| **Handover Failure Probability (HOFP)** | As required to meet Drop Call Rate | Handover failures per handover |
| **Ping-Pong Rate** | Minimize given Drop Rate KPI achieved | Short dwell time handovers/second in connected mode for a given short dwell time |
| **HO Rate** | Minimize given Drop Rate Achieved | Handovers/second in connected mode; this can be influenced by idle-mode re-selection parameters; |
| **Race Zones** | 0 | A race zone has no stable server and every handover has almost zero dwell time. |
| **Edge SINR Loss** | Minimize or <3dB | Difference between Ideal server $5^{th}$%ile SINR and actual server $5^{th}$%ile SINR over all time in connected mode |

**Table 2: Key Control and Optimization Parameters to achieve target KPIs for intra-carrier handover**

| Parameter | Description |
|---|---|
| ***K*** | Layer 3 measurement filter parameter in connected mode; layer 3 filter equation is $F_n = (1-a)F_{n-1} + aM_n$ where $a = (1/2)^{(\frac{K}{4})}$ with sample interval of 200ms |
| ***timeToTrigger*** | Time duration for A3 event entering condition to trigger a measurement report in connected mode (aka *TTT* or $t_T$) |
| $\boldsymbol{O_{ff}}$ | A3 event measurement offset (dB) in connected mode |
| $\boldsymbol{H_{ys}}$ | A3 event measurement hysteresis (dB) in connected mode |
| $\boldsymbol{O_{cp}}$ | A3 event Cell Individual Offset (CIO) scalar in dB for *all* the neighbors in connected mode (scalar) |
| $\boldsymbol{O_{cn}}$ | A3 event Cell Individual Offset (CIO) in dB for *each* of up to 32 neighbors in connected mode (vector). |

## 2 The Handover Race Condition

As we have seen in the previous section, MRO must address the joint optimization of the A3 event triggering parameters across multiple cells to minimize handover failures and unnecessary handovers. Handover races are highly degrading and yet unnecessary and avoidable handovers in the eICIC context. Although a handover failure is intuitive to the and well known to most (dropped calls may result from failed handovers), the race condition in HTN deployments is a subtler condition. To motivate the discussion, Figure 2 shows a typical RSRP profile around a pico with and without bias. The solid curves represent the RSRP across a mobility path through three cells that include two macros (M1 and M2) and a pico (P1). To expand the pico coverage region, bothpico cell and macro cells transmit CIO parameters to their UE's. (The pico CIO and the macro CIO are assumed to be the additive inverse in this example.) In both cases, the CIO parameters raise the RSRP of the pico relative to the macros for processing of mobility events. The range expansion of P1 via RSRP biasing is shown in the dashed line of Figure 2. A UE moving from M1 into the coverage area of P1, applies a bias to raise P1 RSRP relative to M1 RSRP. This bias is set by the RRC layer of M1, and is an element of $O_{cn}$ vector. The result is an early UE handover from M1 to P1. Once this handover occurs, it is necessary that P1 transmit a different $O_{cn}$ vector to UE that raises the P1 RSRP relative to M1 RSRP, to stop the UE from immediately handing back to M1. It is convenient to use the same magnitude of bias with opposite signs in M1 and P1, as this will keep the 2-cell effective hysteresis the same. From the perspective of range extension of P1 relative to macro layer, it might appear that this would be sufficient. However, in the race zone indicated in Figure 2, if the bias from P1 to M2 was zero, it is clear that a UE connected to P1 (with smaller global hysteresis) would generate an A3 event to handover to M2. Once connected to M2, that UE would then handover from M2 to M1, assuming zero bias in the macro layer, and finally back to P1. These are multiple consecutive ping-pongs, and are referred to in this paper as a *race condition*. In other words, the race zone is the area where UEs continue to handover repeatedly from cell to cell.

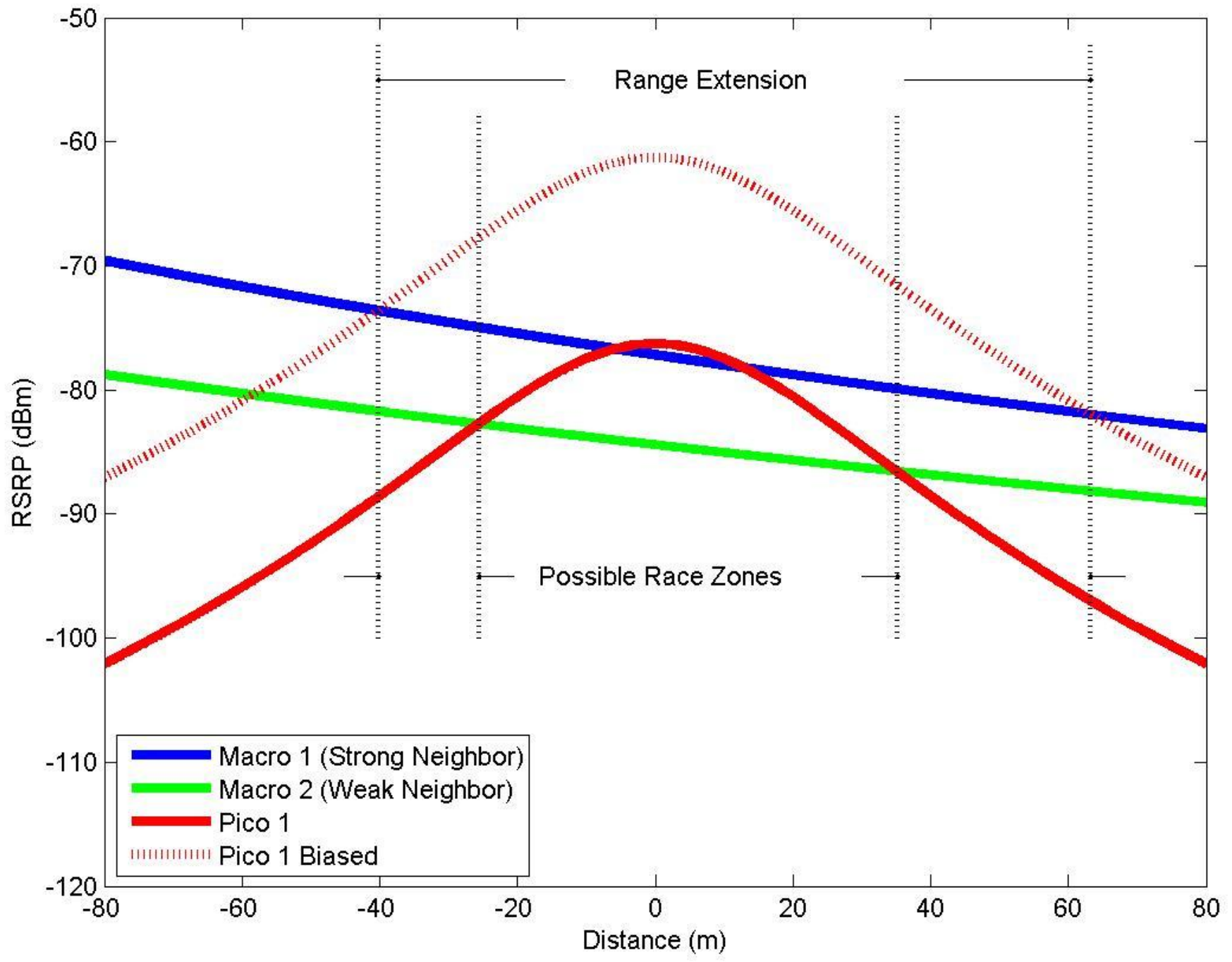

**Figure 2 Illustration of Macro-Macro-Pico Race Condition, and definition of strong neighbor set. Bias matricies that eliminate race condition are shown in Table 3**

A race condition differs from ordinary ping-pongs in that ping-pongs typically involve mobility and its associated multipath fading effects to occur, while a race condition can occur for a perfectly stationary UE. If the UE is non-stationary, the race condition only lasts for the duration of the UE sojourn time in the race zone. For vehicular traffic, race conditions are therefore less problematic unless the race probability is large; however, they can be disruptive for stationary users. There are several good reasons for classifying race as a distinct KPI:

i. A small number of race locations may not appear as significant if aggregated together with ping-pong KPIs but stationary UE's at these locations would be almost in a constant state of handover in connected or re-selection in idle mode, affecting significantly the Quality of Experience (QoE). This problem does not occur with a more general classification of ping-pong, which requires mobility to occur as it is a dynamic issue only.
ii. The stationary UE in a race zone will experience increased battery power consumption with few opportunities to send or receive traffic. The increased handover and re-selection overhead will also increase the MME transactional loading as well as the X2 and S1 interface traffic load.

A handover race is eliminated by optimally constraining the CIO i.e. the effective hysteresis vector $H_{e1} = O_{cp} + O_{ff} - O_{cn} + H_{ys}$. This definition simplifies the A3 condition to, $M_n > M_p + H_{e1}$. The effective hysteresis consists of the *global*[2] *hysteresis* $H_g = O_{ff} + H_{ys}$ ,a quantity that contains all the non-CIO components and the offloading/retaining bias vectors. The difference $O_{cp} - O_{cn}$ is the *retaining bias* for all pico to macro (PM) mobility events, while the difference $O_{cn} - O_{cp}$ is the *off-loading bias* for all macro to pico (MP) mobility events. *In effect, the retaining and off-loading bias are the control knobs to the MRO problem.* As we will explain shortly, increasing retaining bias reduces the race condition probability while increasing offloading bias has the opposite effect.

The CIO assignments can be represented by a bias matrix $B = [b_{ij}]$ where

$$b_{ij} = \begin{cases} O_{cn}(i,j) \text{ if } i \neq j \\ O_{cp}(i) \quad \text{if } i = j \end{cases} \qquad (2)$$

The cell $i$ transmits $O_{cp}(i)$ (typically $O_{cp}(i) = 0$) and $O_{cn}(i,j)$ to the UE . Table 3 shows an example bias matrix that mitigates the race condition above, where each row corresponds to the CIO parameters broadcasted by the P1, M1 and M2 cells in this order.

**Table 3: Bias Matrix for Race Avoidance in Figure 2. Note that there is no benefit to using off-loading bias in Macro 2.**

| | Race Condition | No Race | No Race using Global Retaining Bias[3] |
|---|---|---|---|
| $\boldsymbol{B} = \begin{bmatrix} \text{Pico 1 CIO Vector} \\ \text{Macro 1 CIO Vector} \\ \text{Macro 2 CIO Vector} \end{bmatrix}$ | $B = \begin{bmatrix} 0 & -b & 0 \\ b & 0 & 0 \\ 0 & 0 & 0 \end{bmatrix}$ | $B = \begin{bmatrix} 0 & -b & -b \\ b & 0 & 0 \\ 0 & 0 & 0 \end{bmatrix}$ | $B = \begin{bmatrix} b & 0 & 0 \\ b & 0 & 0 \\ 0 & 0 & 0 \end{bmatrix}$ |

Systematically defining and updating such matrices are the tasks executed SON MRO in LTE eICIC networks.

Two notable side-effects are taking place in this simple example:

- With no bias anywhere, M2 is not a handover target for P1, but after biasing the M1 to P1 handovers, suddenly, M2 has become a handover target for P1, leading to a race condition.

[2] The scope of global-wide parameters is the cluster of cells over which optimization is performed. Clustering methods are outside the scope of this paper.
[3] The role of the global retaining bias will be explained in a subsequent section – it is used when the number of neighbors exceeds 32.

Finally, for the no-race condition in Table 3, M2 is again not a handover target of P1, but it still requires a bias value in P1.

- In this example, only M1 has off-loading benefit. Later we refer to the neighbors with off-loading benefit as strong neighbors. M2 being a weak neighbor to P1 gains no benefit from off-loading bias. Worse, depending on the strength of M2 over P1, M2 may have to assign Almost Blank Sub-frames (ABS), which reduces the available non-ABS resources in M2.

Before proceeding, it is worth addressing another kind of race problem he so-called "pico-pico-macro" (PPM) where picos have pico neighbors that have different optimum bias requirements as shown in Figure 3.

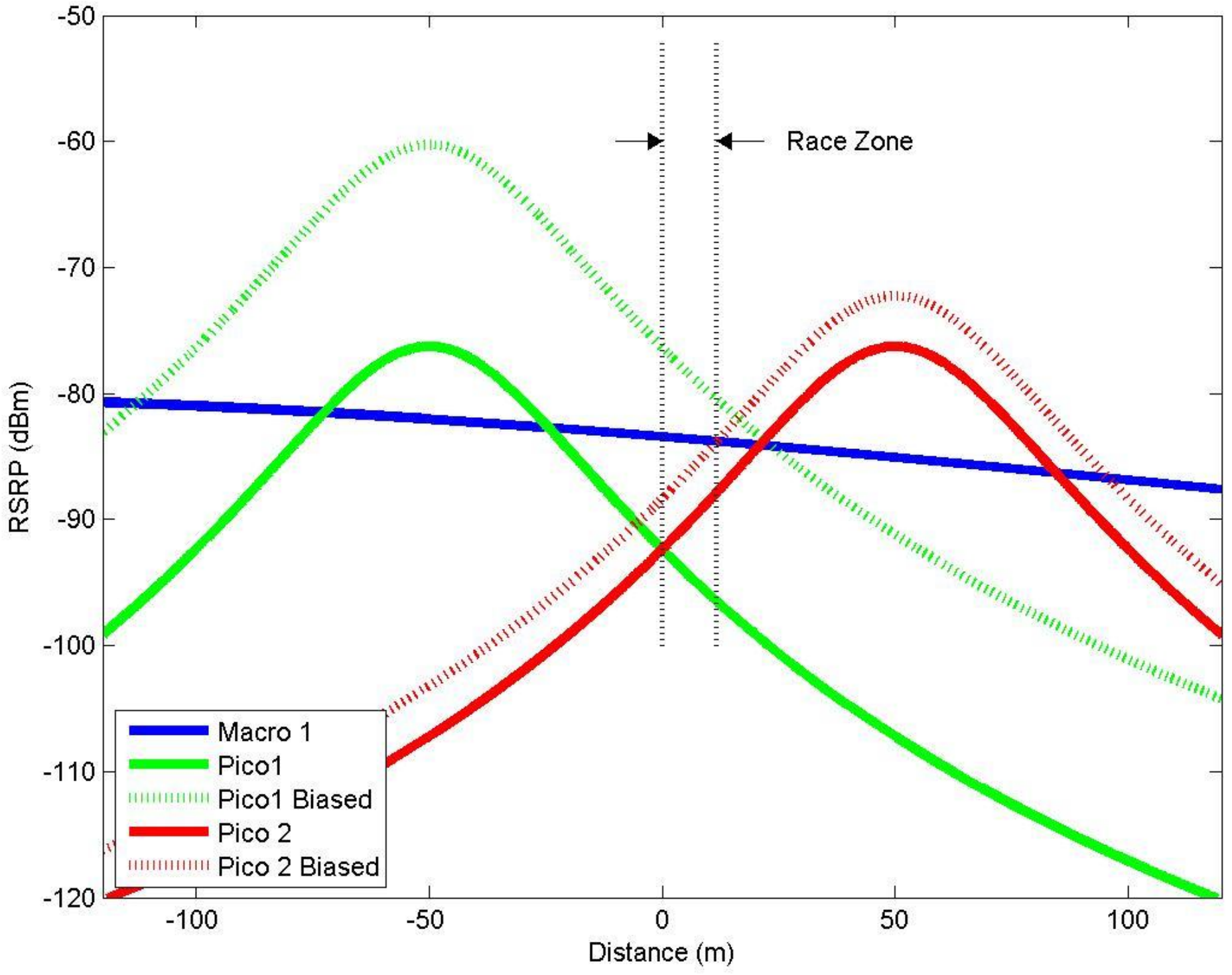

**Figure 3 Illustration of Pico-Pico Race Problem**

The PPM race class is more difficult to solve if an independent range extension bias per pico is maintained. Synchronizing all retaining biases among the pico neighbor groups will solve the problem as detailed in the next section.

# 3 MRO for eICIC HTNs

In this section we introduce the full extent of the MRO eICIC problem and outline possible Bias Matrix assignment solutions. Four options with varying attention to race mitigation are analyzed by simulation to assess the magnitude of the race problem and help decide the best course of action. We start though by recognizing some limitations of the existing eICIC joint ABS-Bias optimization solution [7] designed to maximize network-wide capacity.

With respect to bias, the algorithm in [7] produces one bias per pico which is the retaining and off-loading bias ( $O_{cn}$) to and from all neighboring macros respectively. With respect to ABS patterns, it produces nested[4] ABS patterns, to manifest the fact that pico users, in most cases, receive *strong* interference from *few* neighboring macros. We will call these few strong neighbors as the *off-loading set of neighbors*. When the macro layer uses independent but nested ABS patterns, each UE connected to a pico must be provided with an ABS pattern ideally *common* to all macro interferers.

*However, the macro cell IDs that can potentially interfere across the sojourn time of UEs in a pico CRE region can be very localized due to shadowing effects. The UE may transition in (and out) of such CRE regions, experiencing temporarily higher RSRPs from macros cell IDs that were not considered in the off-loading inducing set of neighbors. If the ABS pattern does not offer protection against all possible interferers, Radio Link Failure (RLF) events can and will happen. Therefore, the number of potential macro interferers that need to be considered can be quite large. The requirement of a common ABS pattern, forces the system to operate in the smallest ABS duty cycle of this larger set, diminishing the eICIC capacity benefits by wasting sub-frames in other macro cells unnecessarily. Detecting such CRE regions where the set of interfering macro cell IDs change, is of fundamental importance to the MRO problem in eICIC HTNs.*

We will call the larger set of interfering neighbors, the *retaining set*. Automatic Neighbor Relation (ANR) SON algorithms can reveal this larger set of interferers – in fact the measurement reports for all neighbors are induced by non-zero bias entries ($O_{cn}$) for up to 32 neighbors - the maximum number allowed by [2]. Ordering the ANR-determined pico neighbors by probability of handover we can assign the $N_{off}$ most probable handover target cell IDs, as members of the off-loading set. The retaining (PMR) set would be the ANR set itself.

It will be seen in the following simulation results that small off-loading sets are sufficient to achieve full range extension and presumably higher capacities. Additionally,the larger the cardinality of the intersection between the offloading and retaining sets, the more pronounced the race conditions and radio link failures can be. We leave the quantification of these tradeoffs for future work.

In conclusion, assignments such as [7] although they maximize network capacity, they can produce radio link failures and race conditions. The solutions described in the next section which may sacrifice some capacity for link quality robustness, are intended to work *in conjunction with* the Joint ABS-Bias Optimization of [7].

---

[4] A nested ABS pattern is a lower duty cycle ABS pattern contained in a higher duty cycle ABS pattern.

## 3.1 CIO Assignment Solutions

### 3.1.1 The Retaining bias (RETB) solution

A common ABS pattern in all macros is assumed but no blanking in any picos is enforced or supported. In RETB, no off-loading bias in neighbor macros is assumed. The pico or a centralized entity can dynamically adjust the retaining bias. There are no race problems due to null off-load bias and the HO rate is small due to large effective hysteresis. Pico-macro (PM) handover failures are reduced as long as bias is not too large or UE are equipped with interference cancellation receivers. Due to lack of off-loading bias, there is a significant reduction in eICIC range extension. In worst case of pico with zero coverage at zero off-loading bias there is no range extension. There is no macro-pico (MP) handover improvement, at least with respect to late HO failures. This solution is intended for multi-vendor networks where the degree of vendor cooperation is small.

### 3.1.2 The Asymmetric Dense-pico deployment (ASYD) solution

In addition to retaining bias as in RETB, the macros use off-loading bias. Whether ABS is dynamic or has a static constant global pattern is independent of the bias management. To minimize neighbor sets, communications, and probability of race, the pico would choose the macros for off-load bias to be no more than the strong macro neighbor set, and all other macro neighbors would use no off-load bias to pico. In addition, this specification is useful when picos have other picos as neighbors with independent bias. This specification increases retaining bias in pico with lower magnitude of optimized bias to value equal to its pico neighbor with largest optimized bias, and the strong macros to reduce off-loading bias to obtain net zero change in load or association level.

There is no PPM or MMP race, and this approach provides effectively close to unconstrained bias optimization of [7] or Joint ABS-Bias Optimization when variable ABS is allowed. When pico density (per macro) becomes large the efficacy of bias as an off-load control is reduced although picos with only pico borders have no coverage dependence on bias. Better solutions might use mean of neighbor group bias $v_j$.

### 3.1.3 Strong retaining neighbors (MINR)

This version has a skew symmetric bias matrix but with strong neighbor, measurement sets for both retaining and off-load. This provides the maximum range extension for given bias without concern for race conditions. Both MMP and PPM races occur in this case. It is cautioned that worse race conditions than this are possible, such as in case where there is retaining and off-load bias only to one strongest macro.

### 3.1.4 Three retaining neighbors (MIN3)

This version has skew symmetric bias matrix with only the best three (in area extension) macro neighbors with off-loading bias to pico. This case explores the sensitivity of range extension and race conditions to a typically small capacity-only driven off-load and retaining bias restriction.

# 4 Performance Analysis of Measurement Neighbor Specifications

In this section we provide performance results that highlight the differences and tradeoffs of the four neighbor specifications (MIN3, MINR, ASYD, and RETB) described in the previous section were tested for race mitigation and range extension.

## 4.1 Methodology and Assumptions

Given the predicted RSRP map and cell locations, the measurement neighbor sets are calculated as defined by the A3 reporting Events controlled by the $O_{cn}$ and $O_{cp}$ bias values that meet a target range extension for each pico and mitigate or eliminate handover race conditions. We start by analyzing the handover at each map grid-point without modeling multipath fading or RRC L3 filtering [2], or without modeling actual UE motion. Leaving out dynamics and ignoring to $t_T$ or $t_f(K)$ does not capture dynamic ping-pong statistics, but it properly captures bias and hysteresis effects that dominate race conditions. This process is used to decide a static definition of explicit neighbor entries in CIO vectors. After this step, a fully dynamic simulation is performed where mobility is explicitly simulated to test for handover failures, handover rate, race, and ping-pong statistics.

Since the exact ABS setting is not important for our purposes, we assume a fixed ABS pattern in any macro that can be involved in a race condition and determine the bias of the pico $v_i$ based on an area equalization objective that divides the original macro coverage area equally between macros and picos. For Release 10 UEs without interference cancellation receivers we assume $0 \leq v_i \leq 5\text{dB}$. This is a moderate upper limit as dropped call rate can increase rapidly for larger values for this receiver type.

In RETB, the optimized bias is increased to partially mitigate for null off-load bias and is used as the retaining bias ( $v_{ri} = v_i$), while the offloading bias is set to zero ($v_{oi} = 0$). In ASYD, the retaining bias for a neighboring group of picos is chosen as the largest of the of the group i.e. $v_{ri} = \max_{i \in group}(v_i)$ , while the offloading bias is set as $v_{oi} = 2v_{ri} - v_i$.

For MINR and MIN3, $v_{ri} = v_{oi} = v_i$

The bias matrix for an example system of 4 cells where with cell 1 and 2 are picos and cells 3 and 4 are macros, would be:

$$B = \begin{bmatrix} 0 & 0 & -v_{r1} & -v_{r1} \\ 0 & 0 & -v_{r2} & -v_{r2} \\ +v_{o1} & +v_{o2} & 0 & 0 \\ +v_{o1} & +v_{o2} & 0 & 0 \end{bmatrix}$$

## 4.2 Results

Given, these settings, an existing macro cell LTE network in New York City (NYC) with added pico cells added along street drive route, is analyzed for bias distribution, area objective attainment, handover race conditions, and measurement neighbor set sizes. The network is exported from an RNP tool to an external handover simulation tool. This NYC cluster includes 152 macro cells and either 0 or 72 pico cells.

The metrics recorded during the evaluations are:

i) Neighbor Lists (retaining and offloading sets) from RSRP map analysis
ii) Race Probability versus Global Hysteresis ($H_g$) from RSRP map analysis
iii) Distribution of $v_{ri}$ and $v_{oi}$ from RSRP map analysis
iv) HO Rate and Dropped Calls from simulation
v) Time-of-Stay Distribution showing evidence of race conditions from simulation
vi) Distributions of pico and residual macro coverage from simulation

The zero pico case provides a macro-only baseline performance. The 72 pico case is evaluated without eICIC and with the four eICIC neighbor set specifications. The simulation has picos deployed with 8m antenna height for selected street segments with approximately 100m inter-pico separation. The street drive route is 47.3km long where the drive route includes every street at least once in both directions. The total length of all unique (counted once) street segments is 21.3km and the length of street segments where picos are deployed is ~7km, so the pico effectiveness in the drive route results may only be ~33% of that ultimately achievable.

Figure 4 shows the handover rate and dropped call rate partitioned into the types of handover for macro-only case, HTN case without eICIC, and four cases of different neighbor specification. The handover types are labeled macro-macro (MM), macro-pico (MP), pico-macro (PM) and Pico-Pico(PP). The drive speed of 60km/h is assumed and the shadow fading is log-normal with 25m correlation distance and 8dB standard deviation. It may be surprising to some that the overall handover rate and dropped call rate both decrease in the HTN deployment versus the macro homogeneous deployment, as these results differ from the results and conclusions in [5], likely due to the following differences in simulation assumptions:

i. Real macro layer deployments in heavily urban areas such as NYC are far from the ideal clover-leaf macro cells layouts in 3GPP models. Simulated outdoor picos are placed along the streets with small inter-pico spacing. In this case, there will be only one near neighbor (another pico) in each direction along the street and interference is less than the typically experienced in a macro-only cloverleaf deployment.
ii. The slope of distance loss (dB/meter) is inversely proportional to cell size, so the area surrounding a pico that is significantly affected by shadowing is much smaller. In other words, distance loss dominates shadow loss more for smaller cells.
iii. As a result of (i) and (ii) the pico-pico handover failure rates are smaller than handover failure rates in a macro only deployment.

In addition to notable HTN improvement in HO rate and dropped calls from macro-only case to HTN, Figure 4 also shows notable improvement with eICIC. Handover race conditions resulting from eICIC are indicated in Time-of-Stay (TOS) distribution in Figure 6 where 2.5% of all UEs have TOS less than 200ms.

Figure 7 shows the RSRP map estimated race probability of ~2.5% with 0dB Hysteresis while at 2dB Hysteresis the race probability drops to ~0.001%. The probability that TOS is less than 500ms for different handover types are given in Table 4. The table data suggests that only MMP handovers are significant in the NYC environment for the MIN3 case. Figure 8 shows that the driving distance per cell type – macros tend to significantly offload mobile UEs to picos even without eICIC.

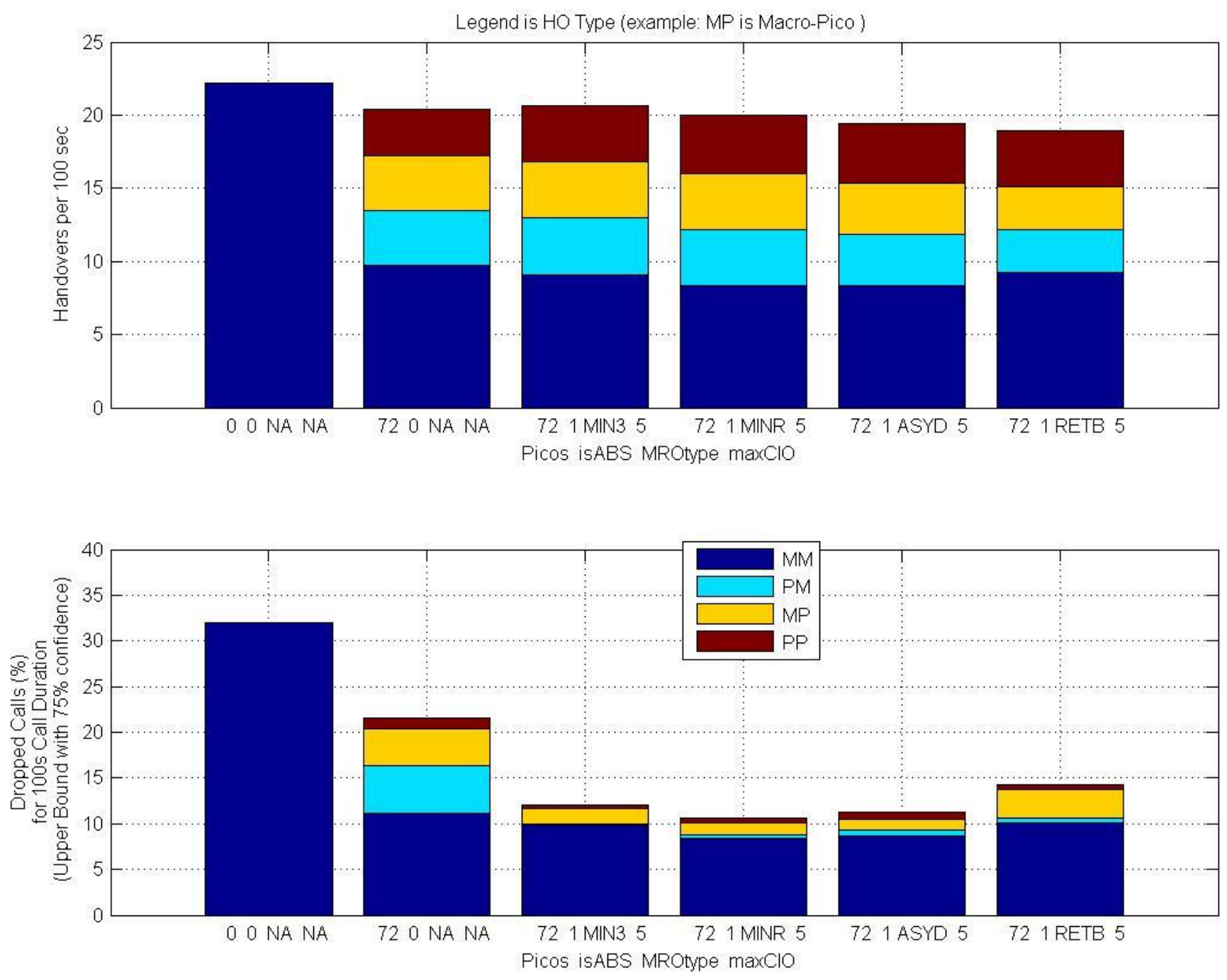

**Figure 4 Handover Rate and Dropped Calls for NYC, Macro-Only system, HTN system, and 4 cases of eICIC HTN system**

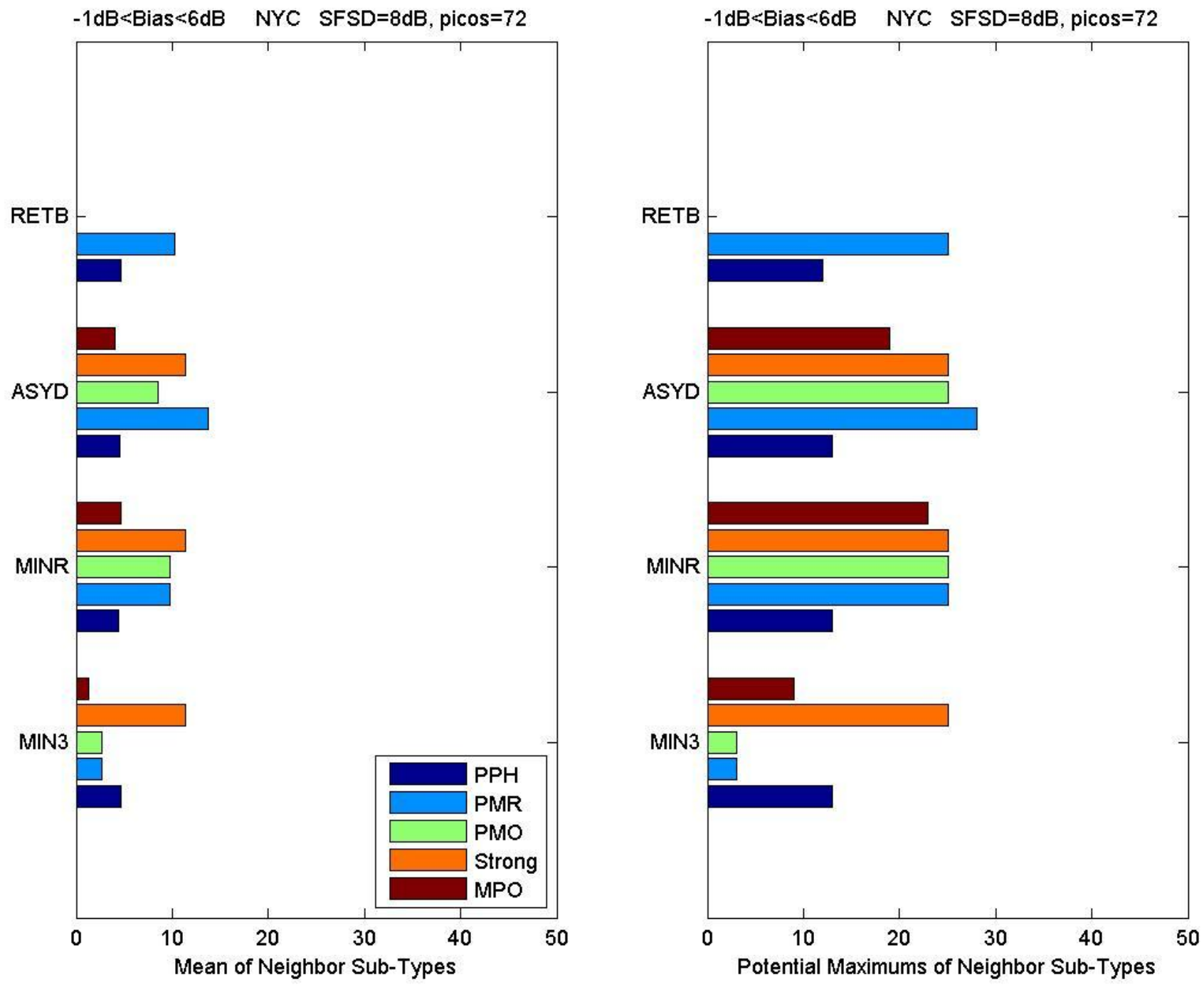

**Figure 5 Neighbor cardinalities derived from RSRP map analysis and used in simulation of NYC environment**

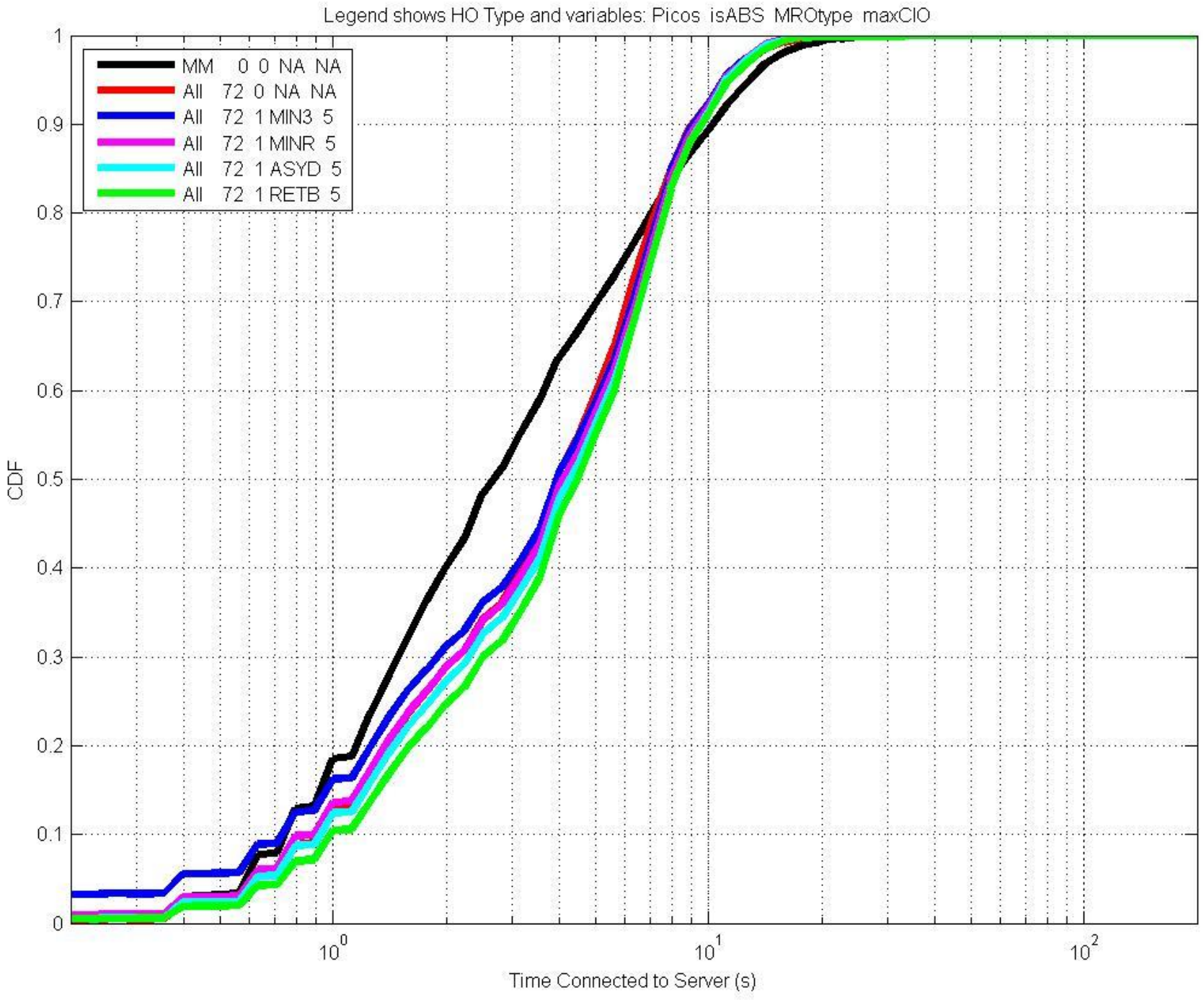

**Figure 6 Time-of-Stay for NYC, Macro-Only System, HTN system, and three cases of eICIC from simulation for NYC**

### Table 4 NYC Environment Simulation Results

| | | | | | | |
|---|---|---|---|---|---|---|
| | Input Parameters | | | | | |
| **Picos** | 0 | 72 | 72 | 72 | 72 | 72 |
| **isABS** | 0 | 0 | 1 | 1 | 1 | 1 |
| **MROtype** | NA | NA | MIN3 | MINR | ASYD | RETB |
| **maxCIO** | NA | NA | 5 | 5 | 5 | 5 |
| | **Results** | | | | | |
| **HO Event Fail Rate (%) All** | 1.4 | 1.01 | 0.56 | 0.51 | 0.55 | 0.72 |
| **HO Event Fail Rate (%) MM** | 1.4 | 1.13 | 1.08 | 1.01 | 1.04 | 1.09 |
| **HO Event Fail Rate (%) PM** | - | 1.41 | 0.03 | 0.10 | 0.19 | 0.17 |
| **HO Event Fail Rate (%) MP** | - | 1.10 | 0.44 | 0.33 | 0.31 | 1.07 |
| **HO Event Fail Rate (%) PP** | - | 0.34 | 0.13 | 0.14 | 0.21 | 0.15 |
| **Prob(100sCallDrop) (%) All** | 27.5 | 18.7 | 11.0 | 9.65 | 10.2 | 12.8 |
| **Prob(100sCallDrop) (%) MM** | 27.5 | 10.5 | 9.38 | 8.12 | 8.34 | 9.67 |
| **Prob(100sCallDrop) (%) PM** | - | 5.15 | 0.11 | 0.40 | 0.67 | 0.49 |
| **Prob(100sCallDrop) (%) MP** | - | 4.04 | 1.69 | 1.27 | 1.10 | 3.09 |
| **Prob(100sCallDrop) (%) PP** | - | 1.10 | 0.49 | 0.58 | 0.84 | 0.58 |
| **HO/100s All** | 22.2 | 20.4 | 20.6 | 20.0 | 19.5 | 18.9 |
| **HO/100s MM** | 22.2 | 9.8 | 9.1 | 8.3 | 8.3 | 9.3 |
| **HO/100s PM** | - | 3.7 | 3.9 | 3.9 | 3.5 | 2.9 |
| **HO/100s MP** | - | 3.7 | 3.9 | 3.8 | 3.5 | 2.9 |
| **HO/100s PP** | - | 3.2 | 3.8 | 3.9 | 4.1 | 3.9 |
| **Prob(T OS<500ms) (%) All** | 3.2 | 2.3 | 5.6 | 3.0 | 2.4 | 1.9 |
| **Prob(T OS<500ms) (%) MM** | 3.2 | 2.7 | 3.8 | 2.6 | 2.5 | 2.6 |
| **Prob(T OS<500ms) (%) PM** | - | 3.8 | 12.4 | 4.5 | 3.3 | 2.5 |
| **Prob(T OS<500ms) (%) MP** | - | 1.4 | 7.2 | 3.4 | 3.0 | 0.6 |
| **Prob(T OS<500ms) (%) PP** | - | 0.5 | 1.4 | 1.8 | 0.8 | 1.0 |
| **Median TOS(s) All** | 3 | 4 | 4 | 4 | 4 | 4 |
| **5thPC TOS(s) All** | 0.6 | 0.6 | 0.4 | 0.6 | 0.6 | 0.7 |
| **Mean TOS(s) All** | 4 | 5 | 5 | 5 | 5 | 5 |

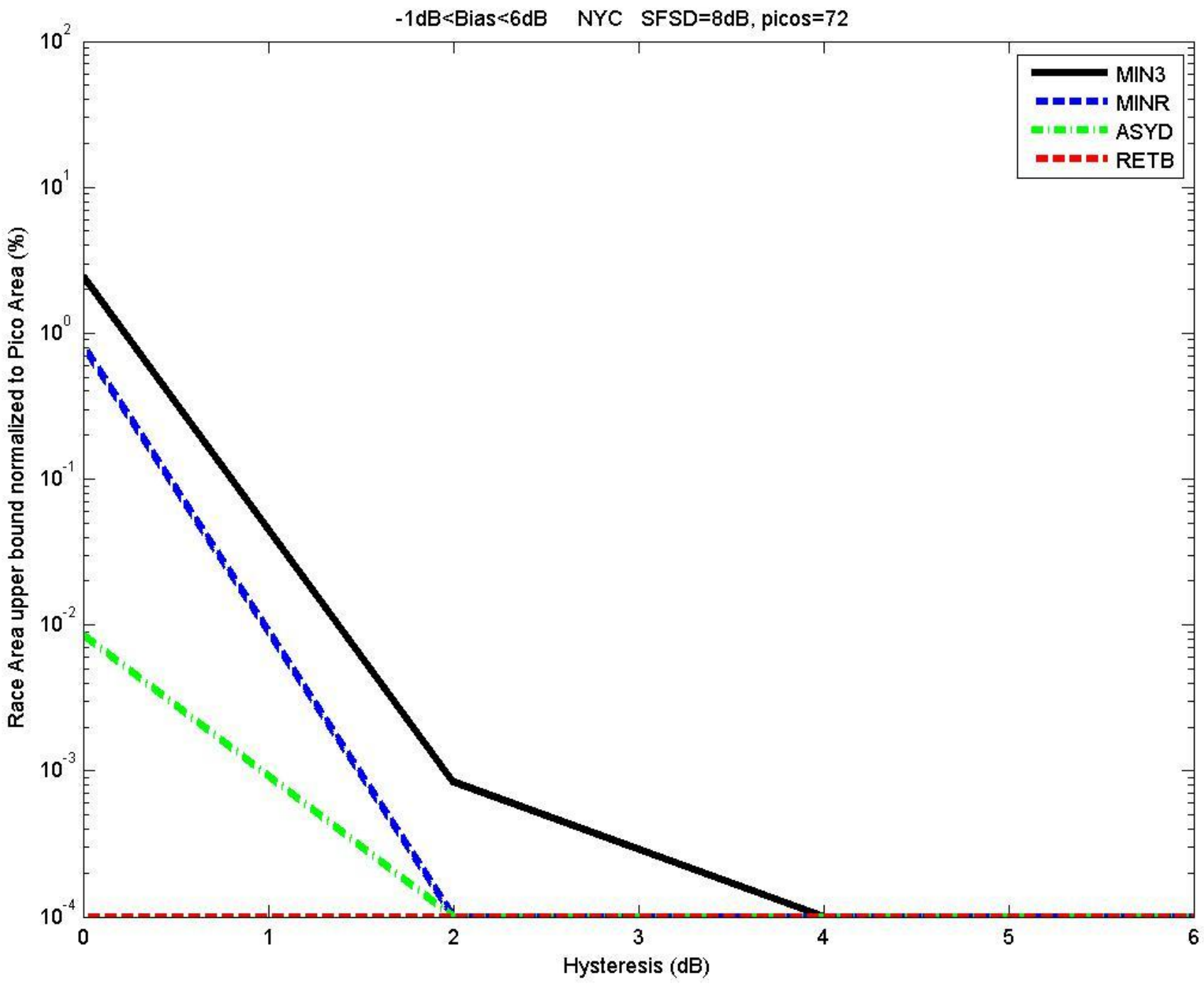

Figure 7 RSRP Map Estimated Race Probability normalized to pico area coverage for NYC

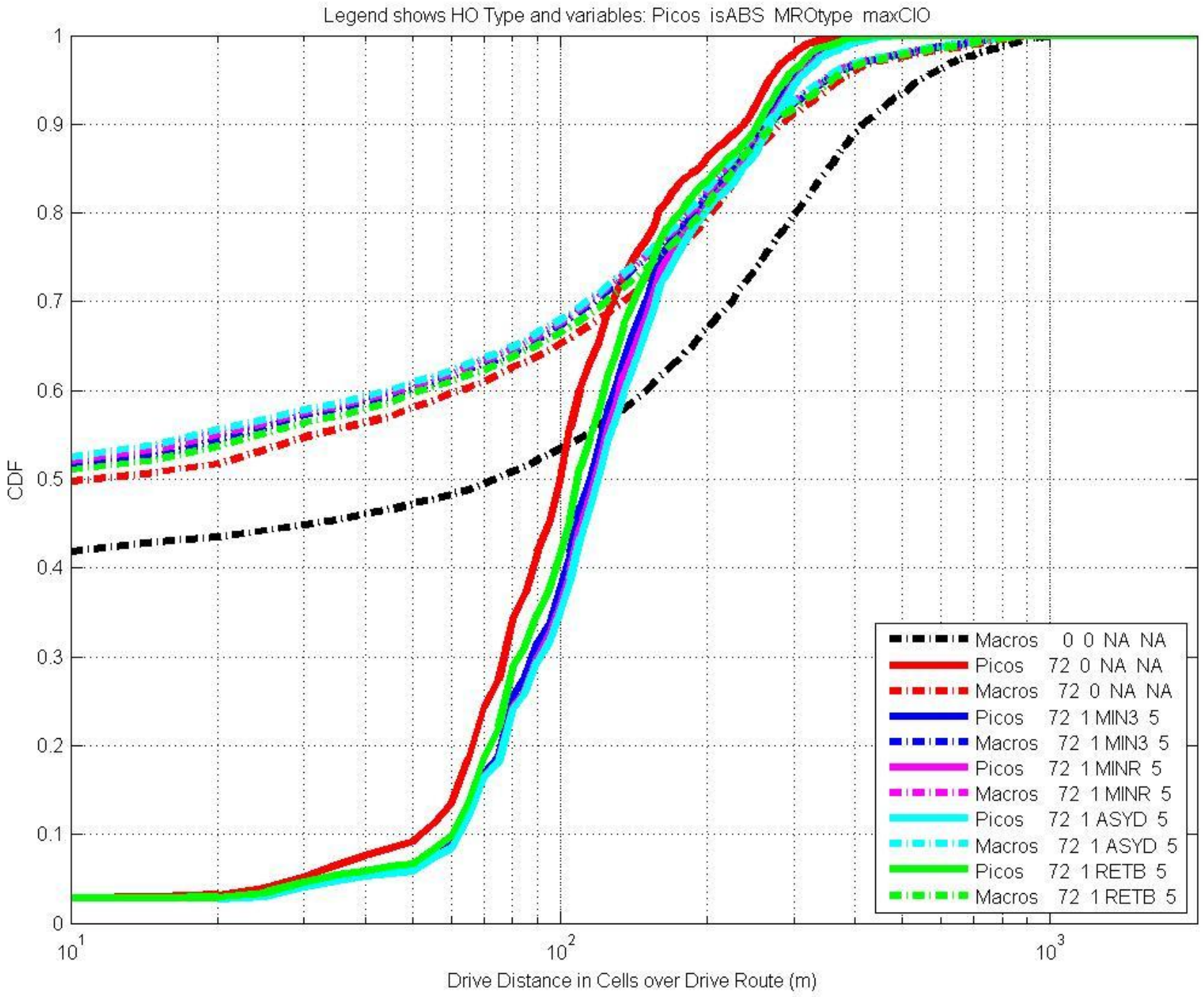

Figure 8 Driving distance per cell type over street drive route showing macro off-load and pico extension

In eICIC, race conditions are just noticeable with MIN3 and MINR.  The race observation of 3% for MIN3 over all handovers has high significance, as these times-of-stay are ~ 900 of ~ 15000 handovers. ASYD is the best of these three in race at the expense of dropped calls. MINR behaves very robustly at least for the bias range considered.

## 5 Conclusions

We presented CIO control techniques for LTE HTNs with eICIC. The results suggest that although a small set of offloading macros is sufficient for capacity optimization, race mitigation requires retaining biases towards a much larger set of macros and a requirement that all retaining set members must offer either a common ABS pattern, or a capability in the RAN to dynamically change each UE ABS/bias configuration. ANR can be used to populate the retaining set and maintaining a per-UE retaining set would allow the dynamic reconfiguration in areas where a much smaller set of interferers prevails therefore avoiding capacity loosing global configurations and maintaining mobility KPIs.

Although not an initial objective of this paper, it is observed that same carrier street level deployment of picos and roof-top deployment of macros has different dropped call and handover rate performance than concluded in [5].

## 6 Acknowledgements

The authors would like to thank Supratim Deb, Chris Mooney for helpful discussions that assisted our understanding of MRO, eICIC and the further development of the techniques of this paper.